\documentclass[10pt]{article} 
\usepackage{amssymb,latexsym, amsmath} 
\usepackage{amscd} 
\newtheorem{example}{\it Example} 
\newtheorem{remark}{\it Remark} 
\newtheorem{theorem}{Theorem} 
\newtheorem{proposition}{Proposition} 
\newtheorem{corollary}{Corollary}

\newtheorem{lemma}{Lemma}

\newcommand{\R}{\mathbb{R}} 
\newcommand{\C}{\mathbb{C}} 
 
\newcommand{\ddim}{\mathrm{ddim\;}} 
\newcommand{\dind}{\mathrm{dind\;}} 
\newcommand{\corank}{\mathrm{corank\;}} 
 
\DeclareMathOperator{\ann}{\mathrm{ann}} 
\newcommand{\ad}{\mathrm{ad}} 
\DeclareMathOperator{\Ad}{\mathrm{Ad}} 
\DeclareMathOperator{\Span}{\mathrm{span\,}}

\DeclareMathOperator{\rank}{\mathrm{rank}} 
 \DeclareMathOperator{\pr}{\mathrm{pr}} 
\begin{document} 
 
\title{Magnetic Flows on Homogeneous Spaces} 
 
\author{Alexey V. Bolsinov \footnote{Supported by RFBR  05-01-00978}
 , \, 
Bo\v zidar Jovanovi\' c  \footnote{Supported by the Serbian Ministry of Science, Project
"Geometry and Topology of Manifolds and Integrable Dynamical Systems".} \\ 
\\ 
\small Dept. of Math. Sciences, Loughborough University\\
\small Loughborough, Leicestershire  LE11 3TU UK, e-mail: A.Bolsinov$@$lboro.ac.uk\\
\small Department of Mechanics and Mathematics, Moscow State University \\ 
\small 119992, Moscow, Russia, e-mail: bolsinov$@$mech.math.msu.su \\
\small and\\ 
\small Mathematical Institute SANU\\ 
\small Kneza Mihaila 35, 11000 Belgrade, Serbia, e-mail: bozaj$@$mi.sanu.ac.yu}
 
%\date{} 
\maketitle 
\begin{abstract} 
We consider magnetic geodesic flows of the normal metrics on a class of
homogeneous spaces, in particular
(co)adjoint orbits of compact 
Lie groups. We give the  
proof of the non-commutative integrability of flows and show, in addition, 
for the case of (co)adjoint orbits,
the usual Liouville integrability by means of  analytic
integrals. We also consider the potential systems on 
adjoint orbits, which are generalizations of the magnetic 
spherical pendulum. 
The complete integrability of such system is proved for an 
arbitrary adjoint orbit of a compact semisimple Lie group.
\end{abstract} 

\centerline{\it to appear in Commentarii Mathematici Helvetici}

\tableofcontents

\section{Introduction} 

Let $Q$ be a smooth manifold with a Riemannian metric $g=(g_{ij})$. 
Consider an arbitrary local coordinate system $x^1,\dots,x^n$ and pass 
from velocities $\dot x^i$ to momenta 
$p_j$ by using the standard transformation $p_j= g_{ij}\dot x^i$. 
Then $x^i, p_i$ 
$(i=1,\dots,n)$ represent the local coordinate system on the cotangent bundle $T^*Q$ 
and the standard symplectic form on $T^*Q$ reads $\omega=\sum 
dp_i\wedge dx^i$. The corresponding canonical Poisson brackets are 
given by 
$$ 
\{f,g\}_0=\sum_{i=1}^{n}\left( \frac{\partial f}{\partial x^i}\frac{\partial g}{\partial p_i} 
-\frac{\partial g}{\partial x^i}\frac{\partial f}{\partial p_i}\right). 
$$ 
 
The equations of the geodesic flow have the Hamiltonian form on $(T^*Q,\omega)$: 
\begin{equation} 
\frac{df}{dt}=\{f,H\}_0 \quad \Longleftrightarrow \quad 
\frac{dx^i}{dt}=\frac{\partial H}{\partial p_i},\qquad 
\frac{dp_i}{dt}=-\frac{\partial H}{\partial x^i}, 
\label{geod_flow} 
\end{equation} 
where the Hamiltonian $H$ is 
\begin{equation} 
H(x,p)=\frac{1}{2}\sum_{i,j=1}^n g^{ij}p_i p_j = \frac{1}{2}\sum_{i,j=1}^n 
g_{ij}\dot x^i \dot x^j. 
\label{Hamiltonian}\end{equation} 
Here $g^{ij}$ are the coefficients of the tensor inverse to the metric. 
 
The geodesic flow can be interpreted as the inertial motion of a particle 
on $Q$ with the kinetic 
energy given by (\ref{Hamiltonian}). 
The motion of the particle under the influence of 
the additional magnetic field given by a closed 
2-form 
$$ 
\Omega=\sum_{1 \le i<j \le n} F_{ij}(x) dx^i \wedge dx^j, 
$$ 
is described by the following equations: 
\begin{equation} 
\frac{dx^i}{dt}=\frac{\partial H}{\partial p_i}, 
\qquad \frac{dp_i}{dt}=-\frac{\partial H}{\partial x^i}+\sum_{j=1}^{n} F_{ij} \frac{\partial H}{\partial p_j}. 
\label{magnetic_flow} 
\end{equation} 
 
The equations (\ref{magnetic_flow}) are Hamiltonian with 
respect to the "twisted" symplectic form $\omega+\rho^*\Omega$, 
where $\rho: T^*Q\to Q$ is the natural projection. 
Namely, the new Poisson bracket is given by 
\begin{equation} 
\{f,g\}=\{f,g\}_0+\sum_{i,j=1}^{n} F_{ij} \frac{\partial f}{\partial p_i}\frac{\partial g}{\partial p_j}, 
\label{magnetic_bracket} 
\end{equation} 
and the Hamiltonian equations $\dot f=\{f,H\}$ read (\ref{magnetic_flow}). 
 
The flow (\ref{magnetic_flow}) is called {\it magnetic geodesic flow} on the Riemannian manifold 
$(Q,g)$ with respect to the magnetic field $\Omega$. 
For simplicity, we shall refer to (\ref{magnetic_bracket}) 
as a {\it magnetic Poisson bracket} and to $(T^*Q,\omega+\rho^*\Omega)$ as a 
{\it magnetic cotangent bundle}.
 
\paragraph{Outline and Results of the Paper.} 
Our work has been inspired
by a recent paper by Efimov \cite{Ef2} in which he proved the
non-commutative integrability of magnetic geodesic flows on
coadjoint orbits of compact Lie groups.
We observed that using the
approach developed in \cite{BJ2}, \cite{BJ3} one can extend this
result to a wider class of homogeneous spaces and
construct, in many cases, complete algebras of commuting integrals.
Besides, this technics turned out to be useful in the theory of integrable
magnetic potential systems. Such systems on Stiefel and Grassmann
manifolds  of two-dimensional planes
in $\mathbb R^n$ and complex projective spaces were studied in \cite{RS, Sak}.

In Section 2, we recall the concept of
non-commutative integrability suggested by Mishchenko and Fomenko
\cite{MF2} and its relation with the Hamiltonian group action established
in \cite{BJ2}.  In Section 3 we introduce a class of homogeneous spaces
$G/H$ admitting a natural $G$-invariant magnetic
field. Briefly, this construction can be explained as follows.
Let $G$ be a compact Lie group, $H$ its closed subgroup and
$\mathfrak h$
and $\mathfrak g$ denote the Lie algebras of  $H$ and $G$, respectively.
Suppose that $a\in\mathfrak h$ is $H$-adjoint invariant. In particular,
$H\subset G_a$, where $G_a \subset G$ is the $G$-adjoint isotropy group of
$a$.  Consider the adjoint orbit $\mathcal O(a)$ through $a$ endowed with
the standard Kirillov-Konstant symplectic form $\Omega_{KK}$ (we can
naturally identify adjoint and coadjoint orbits by the use of
$\Ad_G$-invariant scalar product on $\mathfrak g$).  Then we have the
canonical submersion of homogeneous spaces $ \sigma: G/H \to G/G_a\cong
\mathcal O(a) $ and the closed two-form $\Omega=\sigma^*\Omega_{KK}$ gives
us the required magnetic field on $G/H$.  In particular, for $H=G_a$ we
obtain an adjoint orbit with magnetic term being the Kirillov-Konstant
form.

We prove the non-commutative integrability of
geodesic magnetic flows of the normal metrics (Theorem \ref{main})
on $G/H$ and show, in addition, that for the case
of adjoint orbits one can find
enough commuting analytic integrals (Theorem \ref{commutative}).
The proof is based on recent results concerning geodesic
flows on homogeneous spaces \cite{BJ1, BJ3, MP}.

In Section 4, we study the motion of a particle on coadjoint orbits
under the influence of an additional potential force field.
%The equations of the system are obtained by
%modification of an appropriate Hamiltonian flow without a magnetic term given in \cite{BBC}.  
For the Lie algebra $so(3)$, the system represents
the magnetic spherical pendulum. The generalization of the magnetic
spherical pendulum to the complex projective spaces is obtained in
\cite{Sak}.

In Section 5, we give a representation of the system in the
semi-direct product $\mathfrak g \oplus_\ad \mathfrak g$ and, following
the bi-Hamiltonian approach,
prove its complete integrability for coadjoint orbits of compact
semisimple Lie groups (Corollary \ref{complete1}, Theorem \ref{complete2}).
Various aspects of representations of (polarized) coadjoint orbits in
semi-direct products as  magnetic cotangent bundles
are studied in \cite{NS, KP, DET, RS, Ba}.

Let us emphasize that in the present paper we consider integrability as
a qualitative phenomenon: the phase space of the system is 
foliated almost everywhere by isotropic invariant tori with quasiperiodic dynamics. 
Sometimes this property is not the same as the classical integrability, i.e., 
existence of explicit formulae
for solutions. The simplest example that demonstrates the difference between
two types of integrability is the magnetic geodesic flow on a compact constant
negative curvature surface. We can represent the geodesics explicitly as
projections of magnetic lines from the hyperbolic 2-plane (lines of constant geodesic curvature),
but the restriction of the flow onto energy levels $H > h_{cr}$ are Anosov flows, while
the restriction onto energy levels $H < h_{cr}$ are analytically integrable,
where $h_{cr}$ is some critical level of energy (e.g., see \cite{H, Ta}).

\section{Integrable Systems Related to Hamiltonian Actions}

There are a lot of examples of integrable Hamiltonian systems with 
$n$ degrees of freedom that admit more than $n$ (noncommuting) 
integrals. Then, under some assumptions, the $n$ dimensional Lagrangian 
tori are foliated by lower dimensional isotropic tori. This happens 
in the case of the so-called {\it non-commutative integrability} studied by 
Nekhoroshev \cite{N} and  Mishchenko and Fomenko \cite{MF2} (see also \cite{BJ2, Zu}).

Let $M$ be a Poisson manifold and $\mathcal F$ be a Poisson 
subalgebra of $C^\infty(M)$. Suppose that in the neighborhood of a 
generic point $x$ we can find exactly $l$ independent functions 
$f_1,\dots,f_l \in \mathcal F$ and the corank of the matrix 
$\{f_i,f_j\}$ is equal to some constant $r$. Then numbers $l$ and 
$r$ are called {\it differential dimension} and {\it differential 
index} of $\mathcal F$ and they are denoted by $\ddim\mathcal F$ 
and $\dind\mathcal F$, respectively. The algebra $\mathcal F$ is 
called {\it complete} if: 
$$ 
\ddim{\mathcal F}+\dind{\mathcal F}=\dim M+\corank\{\cdot,\cdot\}, 
$$ 

If $\mathcal F$ is any algebra of functions, then we shall say 
that $\mathcal A\subset \mathcal F$ is {\it a complete subalgebra} 
if 
$$ 
\ddim{\mathcal A}+\dind{\mathcal A}=\ddim{\mathcal 
F}+\dind{\mathcal F}. 
$$ 

The Hamiltonian system  $\dot x=X_H(x)$ is {\it completely 
integrable in the non-commuta\-tive sense} if it possesses a 
complete algebra of first integrals $\mathcal F$. Then (under 
compactness condition) $M$ is almost everywhere foliated by 
$(\dind{\mathcal F}-\corank\{\cdot,\cdot\}$)-dimensional invariant 
isotropic tori. Similarly as in the Liouville theorem, the tori 
are filled up with quasi-periodic trajectories. 
 
Mishchenko and Fomenko stated the conjecture that  non-commutative 
integrable systems $\dot x=X_H(x)$ are integrable in the usual 
commutative sense by means of integrals from $\mathcal A$ that belong 
to the same functional class as the original non-commutative 
algebra of integrals. The conjecture is  proved in 
$C^\infty$--smooth case \cite{BJ2}. In the analytic case, when 
$\mathcal F=\Span_{\mathbb R}\{f_1,\dots,f_l\}$ is a 
finite-dimensional Lie algebra, the conjecture 
 has been proved by Mishchenko and Fomenko in the semisimple case and
just recently by Sadetov \cite{Sa} for  arbitrary Lie algebras.

Now, let a connected compact Lie group $G$ act on a $2n$--dimensional 
connected symplectic manifold $(M,\omega)$. Suppose the action is 
Hamiltonian with the momentum mapping $\Phi: M\to \mathfrak g^* 
\cong \mathfrak g$ ($\mathfrak g^*$ is the dual space of the Lie 
algebra $\mathfrak g$, we use the identification by means of 
$\Ad_G$-invariant scalar product $\langle\cdot,\cdot\rangle$ on 
$\mathfrak g$). 
 
Consider the following two natural classes of functions on $M$. 
Let ${\mathcal F}_1$ be the set of functions in $C^\infty(M)$ 
obtained by pulling-back the algebra $C^\infty(\mathfrak g)$ by 
the moment map ${\mathcal F}_1=\Phi^*C^\infty(\mathfrak g)$. Let 
${\mathcal F}_2$ be  the set of $G$--invariant functions in 
$C^\infty(M)$. The mapping $f\mapsto f\circ\Phi$ is a morphism 
of Poisson structures: 
$$ \{ f\circ\Phi,g \circ\Phi\}(x)= \{f,g\}_{\mathfrak g}(\eta), 
\quad \eta=\Phi(x), $$ 
where $\{\cdot,\cdot\}_{\mathfrak g}$ is the Lie-Poisson bracket 
on $\mathfrak g$: 
$$ 
\{f,g\}_{\mathfrak g}(\eta)=\langle \eta, [\nabla f(\eta),\nabla 
g(\eta)]\rangle, \quad f,g: \mathfrak g\to \Bbb{R}. 
$$ 
Thus, ${\mathcal F}_1$ is closed under the Poisson bracket. Since 
$G$ acts in a Hamiltonian way, ${\mathcal F}_2$ is closed under 
the Poisson bracket as well.  The second essential fact is that 
$h\circ\Phi$ commute with any $G$--invariant function (the Noether 
theorem). In other words: $\{{\mathcal F}_1,{\mathcal F}_2\}=0$. 
 
The following theorem, although it is a reformulation of some well 
known facts about the momentum mapping (e.g., see \cite{GS}), 
is fundamental in the considerations below (see {\rm \cite{BJ2}} for more details).

Let $\mathcal A\subset C^\infty (\mathfrak g)$ be a Lie subalgebra
and $\Phi^*\mathcal A=\{h\circ \Phi, \; h\in \mathcal A\}$ the pull-back 
of $\mathcal A$ by the momentum mapping.
Then we have:

\begin{theorem} \label{Bol_Jov}  
(i) The algebra of functions ${\mathcal F}_1 + {\mathcal F}_2$ is complete: 
$$ 
\ddim({\mathcal F}_1+ {\mathcal F}_2)+ \dind({\mathcal 
F}_1+{\mathcal F}_2)=\dim M. 
$$ 
The dimension of regular invariant isotropic tori, 
common level sets of functions from $\mathcal F_1+\mathcal F_2$, is equal to 
$$
\dim G_\mu-\dim G_x,
$$
for generic $x\in M$, $\mu=\Phi(x)$ ($G_\mu$ and $G_x$ denotes the 
isotropy groups of $G$ action at $\mu$ and $x$). 

(ii) $\Phi^*{\mathcal A}+{\mathcal F}_2$ is a complete algebra on $M$ if 
and only if $\mathcal A$ is a complete algebra on a generic adjoint orbit 
${\mathcal O}(\mu)\subset \Phi(M)$. 
 
(iii) If $\mathcal B$ is complete (commutative) subalgebra of $\mathcal{F}_2$ and 
$\mathcal A$ 
is complete (commutative) algebra on the orbit ${\mathcal 
O}(\mu)$, for generic $\mu \in \Phi(M)$ then $\Phi^*{\mathcal 
A}+{\mathcal B}$ is complete (commutative) algebra on $M$. 
\end{theorem}

 Notice that instead of commutative subalgebras one usually  consider sets of commuting functions.  Clearly,  each commutative set generates a certain commutative subalgebra. The notions of completeness,  $\ddim$ and $\dind$  for a commutative set are defined just in the same way as above.

\section{Magnetic Geodesic Flows} 
 
Let $G$ be a compact connected Lie group with the Lie algebra $\mathfrak g=T_e G$. 
Let us fix some bi-invariant metric $ds^2_0$ on $G$, 
i.e., $\Ad_G$--invariant scalar product $\langle\cdot,\cdot\rangle$ on $\mathfrak g$. 
We can identify $\mathfrak g^*$ and $\mathfrak g$ by $\langle\cdot,\cdot\rangle$. 

Consider an arbitrary homogeneous space $G/H$ of the Lie group $G$.
The metric $ds^2_0$ induces so called {\it normal metric} on $G/H$.
We shall denote the normal metric also by $ds^2_0$.
By the use of $ds^2_0$ we identify $T^*G\cong TG$ and $T^*(G/H)\cong T(G/H)$.
Let $\mathfrak h$ be the Lie algebra of $H$
and $\mathfrak g=\mathfrak h+\mathfrak v$ the orthogonal decomposition.
Then $\mathfrak v$ can be naturally identified with $T_{\pi(e)}(G/H)$
and $T^*_{\pi(e)}(G/H)$, where $\pi: G\to G/H$ is the canonical projection.

\paragraph{Construction of the Magnetic Field.}
We introduce a class of homogeneous spaces $G/H$ having a natural 
construction of the magnetic term, consisting of pairs $(G,H)$,
where $H$ have one-point adjoint orbits.
Let $a\in\mathfrak h$ be the $H$-adjoint
invariant. Then $H$ is a subgroup of the $G$-adjoint isotropy group $G_a$ of $a$. 
The adjoint orbit $\mathcal O(a)$ through $a$ carries
the Kirillov-Konstant symplectic form $\Omega_{KK}$.
Then we have canonical submersion of homogeneous spaces
$$
\sigma: G/H \to G/G_a\cong \mathcal O(a)
$$
and the closed two-form $\Omega=\sigma^*\Omega_{KK}$ gives us the required magnetic field on $G/H$.

The form $\Omega$ is $G$-invariant.
From the definition of $\Omega_{KK}$ (see equation (\ref{Kirillov}))
one can easily prove that at the point $\pi(e)$,
$\Omega$ is given by
\begin{equation} 
\Omega(\xi_1,\xi_2)\vert_{\pi(e)}=
-\langle a, [\xi_1,\xi_2]\rangle, \quad   \xi_1,\xi_2\in\mathfrak v \cong T_{\pi(e)}(G/H)
\label{magnetic_form}
\end{equation} 

Below we give another natural description of the form $\Omega$.

\paragraph{Reduction.}
Consider the right action of the Lie subgroup $H$ to $G$: $(g,h)\mapsto  gh$, $g\in G$, $h\in H$,
and extend it to the 
right Hamiltonian action on $T^*G$. 
After identification $\mathfrak h\cong\mathfrak h^*$, 
we get the momentum mapping 
$$ 
\Psi: T^*G \to \mathfrak h, \quad \Psi(g\cdot \xi)= \pr_{\mathfrak 
h} \xi, \quad \xi\in\mathfrak g. 
$$ 
Here $\pr_\mathfrak h$ denotes the orthogonal projection with 
respect to the invariant scalar product $\langle\cdot,\cdot\rangle$.

It is well known that the symplectic reduced space 
$\Psi^{-1}(0)/H$ is symplectomophic to $(T^*(G/H),\omega)$,
where $\omega$ is the canonical symplectic form on $T^*(G/H)$.
On the other side, the 
reduced spaces $\Psi^{-1}(a)/H_a$, are diffeomorphic to 
the fibre bundles over $T^*(G/H)$ with fibres being
the $H$-adjoint orbit through $a$. In 
particular, if we deal with one-point orbit
$\mathcal O_H(a)=\{a\}$, then 
the reduced space is symplectomorphic to the magnetic cotangent
bundle of $G/H$ (see \cite{AM, RatOrt}). Note that for connected $H$,
$a$ is a $H$-adjoint invariant if and only if $a$ belongs to the center of $\mathfrak h$.

\begin{proposition}\label{reduction} 
Let $a\in\mathfrak h$ be the $H$-adjoint invariant and let $\epsilon$ be  a real parameter.
Then the symplectic reduced space $\Psi^{-1}(\epsilon a)/H$ is 
symplectomorphic to the magnetic cotangent bundle $T^*(G/H)$ 
endowed with the symplectic form $\omega+\epsilon\rho^*\Omega$.
The form $\Omega$ is $G$-invariant and at the point $\pi(e)$ 
is given by (\ref{magnetic_form}).
\end{proposition}

\noindent{\it Proof.} 
We only need to describe the magnetic term for $\epsilon=1$.
Take a principal connection $\alpha$
on the $H$-bundle $G\to G/H$, that is a $\mathfrak h$-valued $1$-form with the
property that the distribution $D=\ker \alpha \subset TG$ ({\it horizontal distribution}) 
is $H$-invariant and transversal to
the orbit of $H$-action. For example we can take $\alpha$ such that $D$ is orthogonal
to $H$-orbits
with respect to the fixed bi-invariant metric.
Then $\alpha_g(X)=\pr_{\mathfrak h} (g^{-1}\cdot X)$ and
$
D_g=g\cdot \mathfrak v.
$
Define the 1-form $\alpha_a=\langle \alpha,a\rangle$ on $G$.
The magnetic form $\Omega$ is the unique 2-form determined
by (see Kummer \cite{Ku}) 
$$
d\alpha_a=\pi^*\Omega.
$$
Since $\alpha$ is a connection and $a$ is $H$-invariant, the form $\Omega$ is well defined. 
Also, since $\pi$ is a submersion, $\Omega$ is closed
but need not be exact.

For a vector $X\in T_g G$ we have the unique decomposition $X=X^h+X^v$, into the horizontal part
$X^h\in g\cdot \mathfrak v$ and the vertical part $X^v\in g\cdot \mathfrak h$. 
The alternative description of the magnetic term is 
\begin{equation}
\Omega_{\pi(g)}(X,Y)=\langle a, \beta_g (X^h,Y^h)\rangle, \quad X, \, Y \in T_{\pi(g)} (G/H)
\label{r1}\end{equation}
where 
$X^h, \, Y^h\in T_g G$ are horizontal lifts of $X$ and $Y$ and $\beta$ is
the curvature of the connection ($\mathfrak h$-valued 2-form
on $G$).

Let $\bar X^h, \bar Y^h$ be arbitrary extensions of $X^h$, $Y^h$ to
horizontal vector fields. Then 
$$
\beta(X^h,Y^h)=-\alpha([\bar X^h,\bar Y^h])
$$
where $[\cdot,\cdot]$ is the commutator of vector fields. Now, by taking
the left-invariant extensions of $X^h$ and $Y^h$ we come to the expression
\begin{equation}
\beta (X^h,Y^h)=-\pr_\mathfrak h [g^{-1}\cdot X^h, g^{-1}\cdot Y^h],
\label{r2}\end{equation}
where $[\cdot,\cdot]$ is the Lie algebra commutator.

The horizontal lifts of $\xi_1,\xi_2\in\mathfrak v \cong T_{\pi(e)}(G/H)$
to $\mathfrak g \cong T_e G$ 
are exactly $\xi_1$ and $\xi_2$ considered as elements of $\mathfrak g$. Whence,
from relations (\ref{r1}) and (\ref{r2}) we get the required expression 
(\ref{magnetic_form}) for the magnetic form $\Omega$ at $\pi(e)$. $\Box$

\medskip

Note that the above magnetic cotangent bundles of the homogeneous spaces
naturally appear in the symplectic induction procedure over a point (see \cite{DET}).

The natural left $G$-action on $T^*G$ commutes with the right $H$-action
and leaves the preimage $\Psi^{-1}(\epsilon a)$ invariant. Thus, from the well known
formula for the momentum mapping of the left $G$-action on $T^*G$ we get

\begin{lemma}
The momentum mapping 
\begin{equation} 
\Phi_\epsilon:\, T^*(G/H)\to \mathfrak g , \label{momentum_map} 
\end{equation} 
of the natural $G$-action on $T^*(G/H)$ with respect to the symplectic 
form $\omega+\epsilon \rho^*\Omega$ is given  by 
\begin{equation*} 
\Phi_\epsilon(g\cdot\eta)=\Ad_g(\eta+\epsilon a), 
\quad \eta \in\mathfrak v, \, g\cdot\eta \in T^*_{\pi(g)}(G/H).
\end{equation*} 
\end{lemma}

\paragraph{Magnetic Geodesic Flows of Normal Metrics.}
The Hamiltonian function of the geodesic flow of the normal metric $ds^2_0$ is 
simply given by 
\begin{equation} 
H_0=\frac 12 \langle \Phi_0,\Phi_0 \rangle. \label{normal} 
\end{equation} 
 
Now, let $\mathcal F_1^\epsilon$ be the algebra of all analytic, 
polynomial in momenta, functions of the form 
$$ 
\mathcal F_1^\epsilon=\{p\circ\Phi_\epsilon, \, p\in \R[\mathfrak 
g]\} 
$$ 
and $\mathcal F_2$ be the algebra of all analytic, polynomial in 
momenta, $G$--invariant functions on $T^*(G/H)$. Then 
$$ 
\{\mathcal F_1^\epsilon,\mathcal F_2\}_\epsilon=0, 
$$ 
where 
$\{\cdot,\cdot\}_\epsilon$ are magnetic Poisson bracket with 
respect to 
$ 
\omega+\epsilon\rho^*\Omega. 
$ 
 
Consider the Hamiltonian $H_\epsilon=\frac12 \langle 
\Phi_\epsilon,\Phi_\epsilon \rangle\in \mathcal F_1^\epsilon$. We have 
\begin{eqnarray*} 
H_\epsilon (g\cdot\eta)&=& \frac12 \langle 
\Ad_g\eta,\Ad_g\eta\rangle+ 
\epsilon\langle \Ad_g \eta,\Ad_g a\rangle+\epsilon^2\frac12\langle \Ad_g a,\Ad_g a \rangle \\ 
&=& H_0(g\cdot\eta)+\epsilon^2\frac12 \langle a,a \rangle=H_0(g\cdot\eta)+const, 
\end{eqnarray*} 
where we used that $\eta\in \mathfrak v$ is orthogonal to 
$a\in\mathfrak h$. Thus, we see that Hamiltonian flows of $H_0$ and 
$H_\epsilon$ coincides:
$
\dot f=\{f,H_0\}_\epsilon=\{f,H_\epsilon\}_\epsilon.
$
Since $H_{\epsilon}$ belongs 
to $\mathcal F_1^\epsilon$ its commutes with $\mathcal F_2$.
On the other side, as a composition of the momentum mapping with an invariant polynomial, the 
function $H_\epsilon$ is also $G$-invariant and commutes with $\mathcal F_1^\epsilon$.
Hence
$
\{H_0,\mathcal F_1^\epsilon+\mathcal F_2\}_\epsilon=0.
$

From the above consideration and Theorem \ref{Bol_Jov} we get the following result.
Let, as before, $G$ be a compact Lie group and $H\subset G$ be a closed subgroup 
such that $\Ad_H a=a$ for a certain element $a\in \mathfrak h\subset \mathfrak g$.

\begin{theorem} \label{main} 
The magnetic geodesic flow of the normal metrics $ds^2_0$ on the 
homogeneous space $G/H$ with respect to the 
closed 2-form $\epsilon\Omega$ given by (\ref{magnetic_form}) 
is completely integrable in 
the non-commutative sense. The complete algebra of first integrals 
is $\mathcal F_1^\epsilon+\mathcal F_2$. 
\end{theorem} 

\begin{remark}{\rm
The magnetic geodesic flow can be seen also as the reduction of the
geodesic flow of the bi-invariant metric $ds^2_0$ from the invariant
subspace
$\Psi^{-1}(\epsilon a)$ to $\Psi^{-1}(\epsilon a)/H \cong T^*(G/H)$.
With the above notation, this means that each magnetic geodesic line on
$G/H$ is the projection of a certain geodesic $\gamma(t) \subset G$
$$
\gamma(t)=g_0 \cdot \exp((\xi+\epsilon a)t), \quad t\in\mathbb R,
$$
where $\xi \in\mathfrak o$ and $g_0$ is the initial position.
In such a way, the integrability of the magnetic geodesic flow can be
also studied from the point of view of the symplectic reduction (see
\cite{Jo, Zu}).
More precisely, the reduction of the normal geodesic system from $T^*G$ to
the Poisson manifold $(T^*G)/H$ is
completely integrable in non-commutative sense (see Zung \cite{Zu}).
Since the symplectic leaves in $(T^*G)/H$ are Marsden-Weinstein reduced
spaces, it appears that the symmetry reduction for a generic value of
the momentum map $\Psi$ yields a system which is integrable in the
non-commutative sense. The interpretation
of these reduced systems in terms of the Yang-Mills analogue of the Lorentz
force is
well-known (e.g., see \cite{GS}). The magnetic bundle $\Psi^{-1}(\epsilon
a)/H$, in general, corresponds to
a singular leaf in $(T^*G)/H$ and in this case some complementary work has to be given to prove the integrability.
}\end{remark}

\paragraph{Magnetic Geodesic Flows on Adjoint Orbits.}
Consider the (co)adjoint action of $G$ and the $G$-orbit $\mathcal O(a)$ 
through an element 
$a\in\mathfrak g$. In what follows, 
we shall use the representation of the adjoint orbit as a homogeneous space 
$G/H$, where $H=G_a$ is the isotropy group of $a$. 
Since $G$ is a compact connected Lie group, $G_a$ is also connected (e.g, see \cite{GS}, page 259). 
We have 
$$ 
\ann(a)=\{\xi\in\mathfrak g, \, [\xi,a]=0\}=T_e G_a.
$$ 

By definition, the Kirillov-Konstant symplectic form $\Omega_{KK}$ on 
$G/G_a$ is a $G$--invariant form, given at the point $\pi(e)\in G/G_a$ 
by 
\begin{equation} 
\Omega_{KK}(\xi_1,\xi_2)\vert_{\pi(e)}=
-\langle a, [\xi_1,\xi_2]\rangle, \quad   \xi_1,\xi_2\in\ann(a)^\perp=[a,\mathfrak g],
\label{Kirillov} 
\end{equation} 
where $\xi_1$, $\xi_2$ are considered as tangent vectors to the orbit at $\pi(e)$. 
It follows from Proposition \ref{reduction} that the Kirillov-Konstant form
can be seen as a magnetic form obtained after right symplectic $G_a$-reduction of $T^*G$ as well.
 
From Theorem \ref{main} we recover the Efimov result \cite{Ef2} (see also \cite{BJ4}):

\begin{corollary}
Let $G$ be a compact Lie group and $a\in \mathfrak g$. The 
magnetic geodesic flows of normal metric $ds^2_0$ on the 
(co)adjoint orbit $\mathcal O(a)=G/G_a$ with respect to the magnetic form
$\epsilon\Omega_{KK}$ is completely integrable in the non-commutative sense. 
\end{corollary} 

\begin{remark}{\rm
For a generic $\eta \in \ann(a)^\perp$ we have equality $ \dim 
G_{\eta+\epsilon a}=\dim G_{\eta}$ for all $\epsilon \in 
\mathbb{R}$ (see \cite{BJ3, MP}). It implies that the dimension of 
the regular invariant tori does not depend on $\epsilon$ and is 
equal to 
\begin{equation}
\dind(\mathcal F_1^\epsilon+\mathcal F_2)=\dim G_{\eta}-\dim (G_a)_\eta, 
\label{delta}
\end{equation}
for a generic element $\eta\in \ann(a)^\perp$ (see \cite{BJ1}).  
Therefore, the influence of the magnetic fields $\epsilon\Omega_{KK}$, 
$\epsilon \in \mathbb{R}$ reflects as a deformation of the 
foliation of the phase space 
$(T^*\mathcal O(a),\omega+\epsilon\rho^*\Omega_{KK})$ by invariant isotropic 
tori. As the magnetic field increases, the magnetic geodesic lines 
become more curved. }\end{remark}

\begin{example}{\rm
On the unit round sphere (see the next section),  
the magnetic geodesic lines are 
circles on the sphere. It can be easily proved that for the motion 
with unit velocity, the radius of the circles is equal to 
$ 
r_\epsilon={\rm arctg}(\frac{1}{\vert\epsilon\vert}). 
$ 
As $\vert\epsilon\vert$ tends to infinity, $r_\epsilon$ tends to 
zero, and as $\epsilon$ tends to zero, then $r_\epsilon$ tends to 
$\frac{\pi}2$. 
}\end{example}
 
\paragraph{Commutative Integrability.}
To prove the commutative integrability, we use the argument 
shift method developed by Mischenko and Fomenko \cite{MF1}
as a generalization of  Manakov's construction \cite{Ma}. 

Let $\mathbb{R}[\mathfrak g]^G$ be the algebra of 
$\Ad_{G}$-invariant polynomials on $\mathfrak g$. Then the 
polynomials 
\begin{equation}
{\mathcal A}_c=\{p(\cdot+\lambda c), \;\lambda\in\mathbb{R},\; p\in 
\mathbb{R}[\mathfrak g]^G\} 
\label{A}
\end{equation}
obtained from the invariants by shifting the argument are in 
involution with respect to the Lie-Poisson bracket \cite{MF1}. 
Furthermore, for {\it every} adjoint orbit 
in $\mathfrak g$, one can find $c\in \mathfrak g$, such that 
${\mathcal A}_{c}$ is a complete commutative set of functions on this orbit. 
For regular orbits it is shown in \cite{MF1}. For singular orbits 
there are several different proofs, see \cite{Mik1, Br3, Bo}. 
Thus, the argument shift method allows us to 
construct a complete commutative subalgebra in $\mathcal F_1^\epsilon$. 
%One can use it to construct 
%such a subalgebra in $\mathcal F_2$ as well \cite{BJ3, MP}. 
 
The $G$-invariant, polynomial in momenta functions on $T^*(G/H)$ 
are in one-to-one correspondence with $\Ad_{H}$-invariant polynomials on 
$\mathfrak v$, via their restrictions to $T^*_\pi(e)(G/H)\cong\mathfrak v$.
%$$ 
%f\in \R[\mathfrak v]^H \longmapsto  \bar f\,: T^*(G/H)\to\R, \quad \bar f(g\cdot \eta)=f(\eta), 
%\quad \eta\in\mathfrak v, 
%$$ 
Within this identification, from
(\ref{magnetic_bracket}), (\ref{magnetic_form}) and Thimm's formula 
for $\epsilon=0$ \cite{Th},
the magnetic Poisson bracket $\{\cdot,\cdot\}_\epsilon$ on $T^*(G/H)$ 
corresponds to the following bracket on $\mathbb{R}[\mathfrak 
v]^H$ 
%(our notation is slightly different from Efimov's \cite{Ef2}) 
\begin{equation} 
\{f,g\}^\epsilon_{\mathfrak v}(\eta)=-\langle \eta+\epsilon a, 
[\nabla f(\eta),\nabla g(\eta) ]\rangle, \quad f,g\in\mathbb{R}[\mathfrak v]^H, \label{b2} 
\end{equation} 
where $\mathbb{R}[\mathfrak v]^H$ denotes the algebra of 
$\Ad_{H}$-invariant polynomials  on $\mathfrak v$. 
 
It is interesting that 
\begin{equation}
\Lambda_{\lambda_1,\lambda_2}=\lambda_1 \{\cdot,\cdot\}^0_\mathfrak v
+\lambda_2\{\cdot,\cdot\}^a_\mathfrak v, \quad \lambda_1^2+\lambda_2^2\ne 0
\label{pencil}
\end{equation}
is a pencil of the compatible Poisson brackets on 
$\mathbb{R}[\mathfrak v]^H$. Here
\begin{equation} 
\{f,g\}^a_{\mathfrak v}(\eta)=-\langle a, 
[\nabla f(\eta),\nabla g(\eta) ]\rangle, \quad  f,g\in\mathbb{R}[\mathfrak v]^H.
\label{b3} 
\end{equation} 

By the use of the pencil (\ref{pencil}) and the completeness criterion derived in \cite{Bo}, 
for the case of adjoint orbits, i.e, when $H=G_a$ and $\mathfrak v=\ann(a)^\perp$
one can conclude that the collection of Casimir functions 
of all the brackets $\Lambda_{1,\lambda}$, $\lambda\in\R$:
\begin{equation} 
{\mathcal B}_a=\{p^\lambda_a(\eta)=p(\eta+\lambda a), \; 
\lambda\in\mathbb{R},\;p\in \mathbb{R}[\mathfrak g]^G,\; \eta\in \ann(a)^\perp\} 
\label{shift} 
\end{equation} 
is a complete commutative subset in $\mathbb{R}[\ann(a)^\perp]^{G_a}$ with respect 
to the canonical bracket $\Lambda_{1,0}$. 
If $a$ is regular in $\mathfrak g$ then $G_a$ is a maximal torus. 
In this case the completeness of $\mathcal B_a$ 
can be easily verified (e.g., see \cite{BJ1, BJ3}). A nontrivial generalization 
to singular orbits of classical groups is done in \cite{BJ3, Bul} 
and by Mykytyuk and Panasyuk for a general case \cite{MP}. 
Namely, it is proved that all (complexified) brackets $\Lambda_{\lambda_1,\lambda_2}$
have the same corank in a generic point $\eta\in\mathfrak \ann(a)^\perp$, equal to (\ref{delta}).
It follows from Theorem 1.1 \cite{Bo} that (\ref{shift}) is a complete commutative algebra with 
respect to each Poisson bracket (\ref{pencil}) as well. 

Whence, according Theorem \ref{Bol_Jov}, we get 
the following statement 

\begin{theorem} \label{commutative} 
The magnetic geodesic flows of the normal metric $ds^2_0$ on 
the orbit $\mathcal O(a)=G/G_a$, with respect to the magnetic field $\epsilon\Omega_{KK}$ 
is completely integrable in the commutative sense, by means of analytic, polynomial in 
momenta first integrals $\Phi^*_\epsilon(\mathcal A_c)+\mathcal B_a$, where $\mathcal A_c$ 
and $\mathcal B_a$ are given by (\ref{A}) and (\ref{shift}), respectively.
\end{theorem} 
 
The commutative integrability of the magnetic flows on the 
complex projective spaces is proved by Efimov in \cite{Ef1}
(since the complex projective spaces are symmetric spaces,
in this case the algebra of $\mathcal F_2$ is commutative).

The integrals $\mathcal B_a$ can be used for
deforming the normal metric to a certain class of $G$-invariant 
metrics on $\mathcal O(a)$ with completely integrable magnetic geodesic 
flows. Theorem \ref{commutative} is announced 
in \cite{BJ4}, where one can find the explicit description of the deformed 
flows within the standard representation of
the orbit $\mathcal O(a)$, as a submanifold of $\mathfrak g$.

\begin{remark}{\rm
If $H$ is a subgroup of the isotropy group $G_a$, then
the rank of the bracket $\Lambda_{0,1}$, in general, is smaller then the rank of
the other brackets from the pencil. Then Theorem 1.1 \cite{Bo}
implies that Casimir functions 
of the brackets $\Lambda_{1,\lambda}$, $\lambda\in \R$ do not form a complete set
with respect to the magnetic bracket (\ref{b2}).
In order to get complete commutative algebra in $(\R[\mathfrak v]^H,\Lambda_{1,\epsilon})$, 
one has to find enought additional commutative 
functions among Casimirs of $\Lambda_{0,1}$.
}\end{remark}

\section{Magnetic Pendulum on Adjoint Orbits} 
 
From now on, we shall consider the orbit 
$\mathcal O(a)$ realised as a submanifold of $\mathfrak g$.
In this representation, the geometry of Hamiltonian flows on $T^*\mathcal O(a)$
is studied by Bloch, Brockett and Crouch \cite{BBC}, while
$G$-invariant magnetic geodesic flows are studied in \cite{BJ4}.

The tangent space at $x=\Ad_g(a)$ is simply 
the orthogonal complement to $\ann(x)$. 
Consider the cotangent bundle $T^*\mathcal O(a)$ as a submanifold of
$\mathfrak g\times \mathfrak g$:
$$
T^*\mathcal O(a)=\{(x,p)\, \vert\, x=\Ad_g(a), p\in\ann(x)^\perp\},
$$
with the paring between $p\in T^*_x \mathcal O(a) \cong \ann(x)^\perp$
and $\eta\in T_x\mathcal O(a)$ given by $p(\eta)=\langle p,\eta\rangle$.
Then the canonical symplectic form $\omega$ on $T^*\mathcal O(a)$
can be seen as a restriction of the canonical linear symplectic
form of the ambient space $\mathfrak g\times\mathfrak g$: 
$\sum_{i=1}^{\dim\mathfrak g} dp_i \wedge dx_i$,
where $p_i$, $x_i$
are coordinates of $p$ and $x$ with respect to some base of $\mathfrak g$.

Let $\xi\in\mathfrak g$ and $x=\Ad_g(a)$. Since 
$\xi_x=\frac{d}{ds}\Ad_{\exp(s\xi)}(x)\vert_{s=0}=[\xi,x]$, 
the momentum mapping of the Hamiltonian $G$-action 
$$
g\cdot (x,p)=(\Ad_g x,\Ad_g p)
$$ 
on $(T^*\mathcal O(a),\omega)$ is given by the relation
$
\langle \Phi_0(x,p),\xi \rangle=\langle p,\xi_x\rangle=\langle p,[\xi,x]\rangle.
$
That is $$
\Phi_0(x,p)=[x,p].
$$
Therefore, the momentum mapping
(\ref{momentum_map}), for $\epsilon \ne 0$, and normal 
metric Hamiltonian read 
\begin{eqnarray*} 
&& \Phi_\epsilon (x,p)=[x,p]+\epsilon x, \\ 
&& H_0(x,p)=\frac12 \langle [x,p],[x,p]\rangle. 
\end{eqnarray*}

\paragraph{Magnetic Spherical Pendulum.} 
As an example, consider the Lie group $SO(3)$. The Lie algebra 
$so(3)$ is isomorphic to the Euclidean space 
$\R^3$ with bracket operation being the standard  vector product. 
The adjoint orbits are spheres $\langle \vec x,\vec x\rangle 
=const.$ Let us consider the unit sphere $S^2$ and its cotangent 
bundle $T^*S^2$ realized as a submanifold of  $\R^6$: 
\begin{equation} 
T^*S^2=\{(\vec x,\vec p)\in \R^6 \,\vert\,  
\phi_1=\langle \vec x,\vec x\rangle =1, \, \phi_2=\langle \vec x,\vec p\rangle =0\}. 
\label{constraints} 
\end{equation} 
%The canonical Poisson bracket $\{\cdot,\cdot\}_0$ on $T^*S^2$ can be described, in 
%redundant variables 
%$(\vec x,\vec p)$ with a use of Dirac construction: 
%$$ 
%\{f, g\}_0 =\{f, g\} 
%+\frac{ \{f,\phi_1 \}\{g,\phi_2\}- \{f,\phi_2\} \{g,\phi_1\} } 
%{ \{\phi_1, \phi_2 \} }, 
%$$ 
%$\{\cdot ,\cdot\}$ being the standard Poisson bracket on ${\mathbb R}^{6}$. 
 
The momentum mapping of the natural $SO(3)$ action on $(T^*S^2,\omega)$ is 
$\Phi_0=\vec x\times \vec p$ and the Hamiltonian of the normal 
metric $ds^2_0$ reduces to $H=\frac12 \langle \vec p, \vec p \rangle$. This 
is the kinetic energy of the unit mass particle motion on the 
sphere. Adding the magnetic term $\epsilon \rho^*\Omega_{KK}$ to 
the canonical form represents the influence of the magnetic 
monopole with the force equal to $\epsilon \vec x\times \vec p$. 
It is well known that two famous integrable potential systems on the sphere remain 
integrable after including the magnetic monopole, namely the spherical pendulum 
and the Neumann system.
Let us consider the spherical pendulum. Then the Hamiltonian 
of the system becomes: 
\begin{equation} 
H=\frac12 \langle \vec p,\vec p\rangle - \kappa\langle \vec b, 
\vec x\rangle, \label{pendulum} 
\end{equation} 
where $\vec b$ is a constant unit vector. The equations of the 
motion of the particle with the energy (\ref{pendulum}) under the 
influence of the magnetic force $\epsilon \vec x\times \vec p$, in 
the redundant variables $(x,p)$ are 
$$ 
\frac{d}{dt}\vec x=\vec p, \quad \frac{d}{dt}\vec p=\kappa \vec 
b+\epsilon \vec x\times \vec p+ \lambda \vec x, 
$$ 
where the reaction force $\lambda \vec x$ is determined from the 
condition that the trajectory $(\vec x(t),\vec p(t))$ satisfies 
the constraints $\phi_1=1$, $\phi_2=0$. The system is completely 
integrable due to the linear first integral 
$f=\langle \vec b, \vec x \times\vec p+\epsilon \vec x \rangle$. 
 
\paragraph{Generalization to Adjoint Orbits.} 
The natural generalization of the magnetic spherical pendulum to 
the orbit $\mathcal O(a)$ is the natural mechanical 
system with the kinetic energy given by the normal metric $ds^2_0$ 
and the potential function $V(x)=-\langle b,x\rangle$, i.e., 
with Hamiltonian 
$$ 
H(x,p)=\frac12 \langle [x,p],[x,p]\rangle- \langle 
b,x\rangle, 
$$ 
under the influence of the magnetic force field given 
by $\epsilon \Omega_{KK}$, where $\Omega_{KK}$ is the standard Kirillov-Kostant form.
Similar systems on the complex projective spaces are studied in \cite{Sak}. 

\begin{proposition} \label{equations} 
The equations of the magnetic pendulum, in redundant 
variables $(x,p)$, are given by 
\begin{eqnarray} 
&& \dot x=[x,[p,x]], \label{xx} \\ 
&& \dot p=[p,[p,x]] +  \epsilon [x,p]+b-\pr_{\ann(x)} b,\label{pp} 
\end{eqnarray} 
 \end{proposition} 
 
\noindent{\it Proof.} 
The equations of the geodesic flow of the normal metric are derived in \cite{BBC} (see also \cite{BJ4}).
The flow is given by the equations: 
\begin{eqnarray} 
&& \dot x=[x,[p,x]],  \label{n_x}\\ 
&& \dot p=[p,[p,x]]. \label{n_p} 
\end{eqnarray} 
 
The magnetic and potential forces have no influence to the equation (\ref{n_x}), 
while the second equation takes the form 
\begin{equation} 
\dot p=[p,[p,x]]+ b  + \Pi +\Lambda, \label{Lambda} 
\end{equation} 
where the Lagrange multiplier $\Lambda \in \ann (x)$ is determined 
from the condition that the trajectory $(x(t),p(t))$ belongs to 
$T^*\mathcal O(a)$.  On the other side, the magnetic force $\Pi$ 
is equal to $\epsilon  [x,p]$. For example, one 
can get this relation considering the magnetic geodesic flow of 
the normal metric (i.e., $V(x)\equiv 0$) and the conservation of the 
shifted momentum map 
\begin{eqnarray*} 
\dot \Phi_\epsilon = & [x,\dot p]+[\dot x,p]+\epsilon \dot x= 
[x,[p,[p,x]] + \Pi+\Lambda]+\\ 
&[[x,[p,x]],p]+\epsilon [x,[p,x]]=[x,\Pi+\epsilon[p,x]]=0. 
\end{eqnarray*} 
 
In order to find $\Lambda$, take the (local)  
base $e_1(\tilde x),\dots,e_r(\tilde x)$, of  $\ann(\tilde x)$ ($[e_i(\tilde 
x),\tilde x]=0$), which is orthonormal at $\tilde x=x$. Then 
$\Lambda=\sum_{i=1}^r \lambda_i e_i(x)$. The Lagrange multipliers 
$\lambda_i$ are determined from the conditions 
\begin{equation} 
\frac d{dt}\langle p,e_i(x)\rangle = \langle \dot p,e_i(x)\rangle+\langle p,\dot e_i(x)\rangle=0, 
\quad i=1,\dots,r \label{lambda}\end{equation} 
 
From the identity $[e_i(\tilde x),\tilde x]\equiv 0$ we have
$[\dot e_i(x),x]+[e_i(x),\dot x]=[\dot e_i(x),x]+[e_i(x),[x,[p,x]]]$.
On the other side, the Jacobi identity gives $[e_i(x),[x,[p,x]]]=[[[p,x],e_i(x)],x]$.
Therefore
\begin{equation} 
\dot e_i(x)+[[p,x],e_i(x)]\in \ann(x), \quad i=1,\dots,r. 
\label{e_i} \end{equation} 
 
Finally, combining (\ref{Lambda}), (\ref{lambda}) and (\ref{e_i}), 
we get $\lambda_i=- \langle b,e_i(x)\rangle$, i.e, 
$$ 
\Lambda= - \pr_{\ann(x)}b. 
$$ 
 
This proves (\ref{pp}).  $\Box$ 
 
\section{Integrability of Magnetic Pendulum} 
 
By the use of the momentum mapping $\Phi_\epsilon$, the equations of 
motion of the magnetic pendulum on the orbit $\mathcal O(a)$,  can be 
rewritten in the symmetric form 
\begin{eqnarray} 
&& \dot x=[\Phi_\epsilon,x], \label{x} \\ 
&& \dot \Phi_\epsilon= [x,b]. \label{phi} \end{eqnarray} 
 
This system can be naturally understood via another representation 
of the magnetic bundle $(T^*\mathcal O(a),\omega+\epsilon \rho^*\Omega_{KK})$, as a {\it coadjoint} orbit 
in the dual space of semidirect product $\mathfrak g \oplus_\ad \mathfrak g$. 
 
\paragraph{Realization of $T^*\mathcal O(a)$ in $(\mathfrak g \oplus_\ad \mathfrak g)^*$.} 
To prove the complete integrability below, we shall 
introduce the semi direct product 
$\mathfrak g \oplus_\ad \mathfrak g$ in a slightly unusual way, by 
the use of contraction of Lie algebras. 

From now on we suppose that $G$ is a compact semisimple Lie group.
Then for $\langle \cdot,\cdot\rangle$ we can take the Killing form multiplied by $-1$.
Let 
$
\mathfrak g^\C=\mathfrak g \otimes \C.
$ 
Then $\mathfrak g^\C$ is a semisimple complex Lie algebra. Denote 
by $\mathfrak g_0$ the real semisimple Lie algebra obtained from 
$\mathfrak g^\C$:
$$
\mathfrak g_0=\mathfrak g \oplus  i \mathfrak g, \quad i^2=-1.
$$

Then $\dim \mathfrak g_0=\dim_\R \mathfrak 
g^\C=2\dim_\C \mathfrak g^\C=2\dim \mathfrak g$ and 
$$ 
\rank_\C \mathfrak g^\C=\rank \mathfrak g=r, 
\qquad \rank \mathfrak g_0 = 2\rank \mathfrak g=2r. 
$$ 

Let $p_1,\dots,p_r$ be the set of  basic homogeneous 
invariant polynomials on $\mathfrak g$ considered as complex 
invariant polynomials on $\mathfrak g^\C$. Then their real and 
imaginary parts form a set of basic  polynomial invariants on $\mathfrak 
g_0$. 
 
The real algebra $\mathfrak g_0$ has the symmetric pair 
decomposition: $\mathfrak g_0=\mathfrak g + i\mathfrak g$: 
$$ 
[\mathfrak g,\mathfrak g]\subset \mathfrak g, \quad [\mathfrak g,i\mathfrak g]\subset i\mathfrak g, \quad 
[i\mathfrak g,i\mathfrak g]\subset\mathfrak g, 
$$ 
and one can consider the contraction of $\mathfrak g_0$: the real 
Lie algebra  $\mathfrak g_\theta$ with the same linear space as 
$\mathfrak g_0$ and the Lie bracket defined by 
$$ [\xi_1+i\eta_1,\xi_2+i\eta_2]_\theta=[\xi_1,\xi_2]+i[\eta_1,\xi_2]+i[\xi_1,\eta_2], \quad \xi_i,\eta_i\in \mathfrak g. $$ 
It is clear that $\mathfrak g_\theta$ is 
the semidirect product $\mathfrak g \oplus_\ad i \mathfrak g$,
where the second term $i\mathfrak g$ is considered as a commutative subalgebra.
 
Now, identify 
$\mathfrak g_\theta^*$ with $\mathfrak g_\theta$ and 
$\mathfrak g_0^*$ with $\mathfrak g_0$ by means of nondegenerate 
scalar product 
\begin{equation}
(\xi_1+i\eta_1,\xi_2+i\eta_2)=\langle \xi_1,\xi_2\rangle - \langle \eta_1,\eta_2 \rangle,
\label{scalar_product}
\end{equation}
which is propontional to the Killing form of $\mathfrak g_0$. 
Then the differential of a smooth function $f$ on $\mathfrak g_0$ (or $\mathfrak g_\theta$)
is $\nabla f\vert_{\xi+i\eta}=\nabla_\xi f-i\nabla_\eta f$
and the Lie-Poisson brackets on $\mathfrak g_0$ and $\mathfrak g_\theta$ become
\begin{eqnarray*} 
\{f,g\}_{\mathfrak g_0}(\xi+i\eta) &=&(\xi+i\eta, [\nabla_\xi f - i \nabla_\eta f, 
\nabla_\xi g - i\nabla_\eta g])\\ 
&=& \langle \xi, [\nabla_\xi f,\nabla_\xi g]-[\nabla_\eta f,\nabla_\eta g]\rangle +
\langle \eta, [\nabla_\xi f,\nabla_\eta g] + [\nabla_\eta f,\nabla_\xi g]\rangle, \\
\{f,g\}_{\mathfrak g_\theta}(\xi+i\eta) &=&(\xi+i\eta, [\nabla_\xi f - i \nabla_\eta f, 
\nabla_\xi g - i\nabla_\eta g]_\theta)\\ 
&=& \langle \xi, [\nabla_\xi f,\nabla_\xi g]\rangle +
\langle \eta, [\nabla_\xi f,\nabla_\eta g] + [\nabla_\eta f,\nabla_\xi g]\rangle. 
\end{eqnarray*} 
 
Note that a generic symplectic leaf (coadjoint orbit) in 
$(\mathfrak g_\theta, \{\cdot,\cdot\}_{\mathfrak g_\theta})$ has the same 
dimension as the orbit in $\mathfrak g_0$, that is $2\dim 
\mathfrak g- 2r$ (see \cite{Br}). 

It is well known that the cotangent bundle to the orbit
of the linear representation of a Lie group to the vector space
can be seen as a coadjoint orbit in the dual space of 
the semidirect of the group with the vector space (e.g., see \cite{GS}). 
The similar statement holds for the magnetic
cotangent bundles:

\begin{proposition}\label{semi_direct} 
The mapping 
$ \Theta_\epsilon: T^*\mathcal O(a) \to (\mathfrak g \oplus_\ad i \mathfrak g)^*$
given by 
\begin{equation}
\Theta_\epsilon(x,p)=\Phi_\epsilon(x,p)+i x \label{theta}
\end{equation}
is a symplectomorphism between $T^*\mathcal O(a)$ endowed with the 
twisted symplectic form $\omega+\epsilon\rho^*\Omega_{KK}$ and the 
coadjoint orbit of the element $\epsilon a+i a$ in 
$(\mathfrak g \oplus_\ad i \mathfrak g)^*$ endowed 
with the canonical Kirillov-Konstant symplectic form. 
\end{proposition} 
 
Novikov and Schmeltzer have constucted such mapping for
the orbits of coadjoint representation of the three-dimensional Euclidean 
space motion group which are symplectomorphic to the megnetic cotangent 
bundles of the sphere (see \cite{NS}). This problem is further developed
in \cite{KP, DET, RS, Ba}.

\medskip

\noindent{\it Proof.} 
We use the following general statement. Suppose
that we have a symplctic manifold $M$ endowed with a transitive Hamiltonian
action of a certain Lie group $K$. Consider the corresponding momentum
mapping $\Theta: M \to \mathfrak k^*$. Then from the standard properties of
a momentum mapping it follows that

1) the image of $\Theta$ is a single coadjoint orbit $O\subset \mathfrak k^*$,

2) $\Theta: M\to O$ is a symplectic covering.

%In order for $\Theta$ to be a global symplectomorphism we only need to
%verify that $\Theta$ is one-to-one.

In our case we just need to describe the transitive Hamiltonian action of
the semidirect product $G\times_{\Ad} i\mathfrak g$ on the cotangent bundle
$T^* \mathcal O(a)$. Such an action exists and is very natural.
Indeed, let us consider first the standard action of $G$ on the cotangent
bundle $T^*O(a)$. As we saw above, the corresponding momentum mapping is
exactly $\Phi_\epsilon$. Now we extend this action by adding the
following action of the vector space $\mathfrak g$:
\begin{equation}
\eta\cdot (x,p) = (x, p+\pr_{\ann(x)^\bot}(\eta)), \qquad \eta\in \mathfrak g,
\label{1}
\end{equation}
where $x\in \mathcal O$, $p\in T^*_x \mathcal O(a)=\ann (x)^\bot$.
% and $\pi_x: \mathfrak g \to (\ann x)^\bot$ is the orthogonal projection.
Whence the action of the whole semidirect
product $G\times_{\Ad} i\mathfrak g$ on $T^*\mathcal O(a)$ is given by:
$$
(g,i\eta)\cdot (x,p) = (\Ad_g x, \Ad_g p+\pr_{\ann(\Ad_g x)^\bot}(\eta)), \quad
(g,i\eta) \in G\times_{\Ad}i\mathfrak g.
$$
%Here the pair $(g,\eta)$ is considered as an element from $G\times_{\Ad}i\mathfrak g$.

It is easy to verify that this formula defines an action and this action is
Hamiltonian. We know this for the first component, and the action of the
second one (\ref{1}) is generated by translations along the Hamiltonian vector
fields of the functions 
$$
H_\eta(x)=-\langle x, \eta \rangle. 
$$
(Notice that
this flow is the same for all structures $\omega+\epsilon\rho^*\Omega_{KK}$,
$\epsilon\in \Bbb R$).

Thus the Hamiltonian with respect to $\omega+\epsilon\rho^*\Omega_{KK}$
corresponding to an element $\xi+i\eta\in \mathfrak g \oplus_{\ad} i\mathfrak g$ takes
the form:
\begin{eqnarray*}
H_{\xi+i\eta} (x,p) &=& H_\xi(x,p) + H_\eta(x,p) \\
&=& \langle \Phi_\epsilon (x,p), \xi\rangle - \langle x, \eta \rangle = 
(\Phi_\epsilon (x,p) + ix, \xi + i\eta)
\end{eqnarray*}
This implies that the formula for the momentum mapping
is given by (\ref{theta}),
where we use the natural identification of $(\mathfrak g \oplus_{\ad} i\mathfrak g)^*$
and $\mathfrak g \oplus_{\ad} i\mathfrak g$ by means of the scalar product
(\ref{scalar_product}).

Since the above action is obviously transitive, 
we conclude that $\Theta_\epsilon$ is a symplectic covering over a certain coadjoint orbit. It is
not hard to verify that $\Theta_\epsilon$ is one-to-one with the image, and
therefore is a global symplectomorphism, as required.
$\Box$ 

\paragraph{Compatible Poisson Brackets and Integrability.} Since the Hamiltonian flow 
$$
\dot f=\{f,h\}_{\mathfrak g_\theta}
$$
of 
$h(\xi+i\eta)=\frac12 \langle \xi,\xi\rangle -  \langle b,\eta\rangle$ 
is given by 
\begin{equation} \frac{d}{dt}(\xi+i\eta)= [\eta,b]+i[\xi,\eta], \label{xi_eta} \end{equation} 
from Proposition \ref{semi_direct} we reobtain the equations 
(\ref{x}), (\ref{phi}). 
 
The flow  (\ref{xi_eta}) is completely integrable. This is related 
to the general construction of integrable systems by the use of 
symmetric pair decompositions of Lie algebras and compatibility of 
Poisson brackets $\{\cdot,\cdot\}_{\mathfrak g_0}$, 
$\{\cdot,\cdot\}_{\mathfrak g_\theta}$ and $\{\cdot,\cdot\}_{ib}$, where 
\begin{eqnarray*}
\{f,g\}_{ib}(\xi+i\eta) &=&(ib, [\nabla_\xi f - i \nabla_\eta f, 
\nabla_\xi g - i\nabla_\eta g]) \\
&=&\langle b, [\nabla_\xi f,\nabla_\eta g] + [\nabla_\eta f,\nabla_\xi g]\rangle
\end{eqnarray*}
(Reyman \cite{R}, see also \cite{Bo}). 
 
Let $\mathfrak A$ be the algebra of linear functions on $\ann(b)$, 
lifted to the linear functions on $\mathfrak g_\theta$: 
$$
\mathfrak A=\{ f_\mu(\xi+i \eta)=\langle \mu,\xi\rangle, \, \mu\in\ann(b)\} 
$$ 
and let 
\begin{equation} 
\mathfrak B= \{\mathfrak{Re}(p_j(\lambda 
\xi+i(\eta+\lambda^2 b)),\, \mathfrak{Im}(p_j(\lambda 
\xi+i(\eta+\lambda^2 b)), 
 \, \lambda\in\R,\, j=1,\dots,r\}, 
\label{B}
\end{equation}
where $p_j$ are basic invariant polynomials of the Lie algebra $\mathfrak g^\C$. 
Then $\mathfrak B$ is  commutative, and  $\mathfrak A+\mathfrak B$ is a complete (non-commutative) 
set of integrals of (\ref{xi_eta}) (see \cite{Bo1} or Theorem 1.5 in \cite{Bo}). 
 
From Proposition \ref{semi_direct} we get 
 
\begin{corollary} \label{complete1} 
The magnetic pendulum system (\ref{xx}), (\ref{pp}) is completely integrable
on a generic orbit $\mathcal O(a)$. The 
complete set of integrals is 
$$ 
\Theta_\epsilon^*(\mathfrak A+ \mathfrak B)=\{f(\Phi_\epsilon(x,p)+i x), \, f\in \mathfrak A+\mathfrak B\}. 
$$ 
\end{corollary} 
 
Note that we can always construct a complete commutative subalgebra in $\mathfrak A$  (more precisely, in the symmetric algebra of $\mathfrak A$, i.e. the polynomial algebra generated by linear functions $f_\mu \in {\mathfrak A}$).  Thus we have 
complete commutative integrability. In particular, if $b$ is a 
regular element of $\mathfrak g$, then $\mathfrak A$ is commutative and contained
in $\mathfrak B$. 
 
The integrability of the system (\ref{xi_eta}) on the whole phase 
space $\mathfrak g_\theta$ do not implies  directly the integrability on singular orbits.
As usual (see, for instance \cite{Mik1, Br3, Bo}), to prove
the completeness of integrals on singular orbits, some additional analysis has to be done.
 
Let $\mathcal O(a)$ be an arbitrary orbit.

\begin{theorem} \label{complete2} 
For a generic regular element $b\in\mathfrak g$, the magnetic pendulum system
on $(T^*\mathcal O(a),\omega+\epsilon\rho^*\Omega_{KK})$,
described by equations (\ref{xx}), (\ref{pp}), is completely integrable. The 
complete commutative set of integrals is 
$$ 
\Theta_\epsilon^*\mathfrak B=
\{\mathfrak{Re}(p_j(\lambda 
[x,p]+\epsilon x+i(x+\lambda^2 b)),\, \mathfrak{Im}(p_j(\lambda 
[x,p]+\epsilon x+i(x+\lambda^2 b))\}. 
$$
%In particular, we can take $b$ to be a regular element in $\ann(a)$.
\end{theorem}

\noindent{\it Proof.} 
 $\Theta_\epsilon^*\mathfrak B$ is complete 
on $(T^*\mathcal O(a),\omega+\epsilon\rho^*\Omega_{KK})$
if and only if $\mathfrak B$ is complete on the coadjoint orbit 
$\Theta_\epsilon(T^*\mathcal O(a))$.

Consider the pencil of compatible Poisson structures
$$
\Lambda_{\lambda_1,\lambda_2}=
\lambda_1 \{\cdot,\cdot\}_{\mathfrak g_\theta}+
\lambda_2 \left(\{\cdot,\cdot\}_{\mathfrak g_0}+\{\cdot,\cdot\}_{ib}\right), \quad
\lambda_1,\lambda_2\in {\mathbb{R}}, \;\lambda_1^2+\lambda_2^2\ne 0.
$$

The functions (\ref{B}) are Casimir functions for $\Lambda_{\lambda_1,\lambda_2}$,
where $\lambda_1+\lambda_2\ne 0$, $\lambda_2\ne 0$,
$\lambda=\sqrt{\lambda_2/(\lambda_2+\lambda_1)}$ \cite{R}.
Since we deal with analytic functions, we only need to prove
the completeness of $\mathcal B$ at one point in 
$\Theta_\epsilon(T^*\mathcal O(a))$. Also, 
the completeness of $\mathcal B$ for one element $b$
implies the completeness for a generic $b\in\mathfrak g$.

Consider the point $\epsilon a+ia \in \Theta_\epsilon(T^*\mathcal O(a))$ and take 
an arbitrary regular element $b\in\ann(a)$. 
According Theorem 1.1 \cite{Bo}, $\mathcal B$ is complete at $\epsilon a+ia$
with respect to the Poisson bracket 
$\Lambda_{1,0}=\{\cdot,\cdot\}_{\mathfrak g_\theta}$ if and only if
\begin{eqnarray*}
\mathrm{(A1)}&&\quad  \rank \Lambda_{\lambda_1,\lambda_2}=2\dim\mathfrak g-2r,  \quad 
\mathrm{for\,\, all} 
\quad (\lambda_1,\lambda_2)\ne(1,0).\\
\mathrm{(A2)}&& \quad \dim\{\xi+i\eta\in\ker\Lambda_{1,0}\, 
\vert\, \Lambda_{0,1}(\xi+i\eta,\ker\Lambda_{1,0})=0\}=2r.
\end{eqnarray*}
Here the Poisson brackets $\Lambda_{\lambda_1,\lambda_2}$ are taken
at the point $\epsilon a+ia$ and they 
are considered as skew-symmetric bilinear forms on $\mathfrak g_\theta$.
Furthermore, all objects are assumed to be (again) complexified.

Since $b$ is a regular element in $\mathfrak g$, it follows that 
$ib$ is a regular element of the semisimple Lie algebra $\mathfrak g_0$:
$$
\ann_{\mathfrak g_0}(ib)=\ann_\mathfrak g (b)+i\ann_\mathfrak g (b).
$$
($\ann_\mathfrak g(b)$ is a maximal commutative subalgebra of $\mathfrak g$.)

The condition (A1), for $\lambda_1+\lambda_2\ne 0$ is equivalent to the regularity
of the elements $\lambda\epsilon a+ia+\lambda^2 ib$, $\lambda\in\mathbb{C}$, $\lambda\ne 0$,
in the Lie algebra $\mathfrak g_0$. This follows easily from the regularity of $ib$.

Now, consider the skew-symmetric form $\Lambda_{-1,1}$. We have
\begin{eqnarray*}
\Lambda_{-1,1}(\xi_1-i\eta_1,\xi_2-i\eta_2)&=&
-(\epsilon a+ia,[\xi_1-i\eta_1,\xi_2-i\eta_2]_\theta)\\
&&+(\epsilon a+ia, [\xi_1-i\eta_1,\xi_2-i\eta_2])\\
&&+(ib,[\xi_1-i\eta_1,\xi_2-i\eta_2])\\
&=& -\langle \epsilon a ,[\eta_1,\eta_2]\rangle + \langle b, [\xi_1,\eta_2]+[\eta_1,\xi_2]\rangle
\end{eqnarray*}
Therefore, $\xi-i\eta\in\ker\Lambda_{-1,1}$ if and only if
$$
[b,\eta]=0, \qquad [\xi,b]-[\eta,\epsilon a]=0.
$$
The first equation yields $\eta\in\ann_\mathfrak g(b)$. On the other hand, since 
$b\in\ann_\mathfrak g(a)$, we have $[\ann_\mathfrak g(b),a]=0$.
Thus, the second equation reduces to $[\xi,b]=0$, i.e., $\xi\in\ann_\mathfrak g(b)$
and
$$
\dim\ker\Lambda_{-1,1}=2r.
$$

It remains to verify (A2). 
We have \begin{eqnarray*}
\Lambda_{1,0}(\xi_1-i\eta_1,\xi_2-i\eta_2)&=&
(\epsilon a+ia, [\xi_1-i\eta_1,\xi_2-i\eta_2]_\theta)\\
&=& \langle \epsilon a ,[\xi_1,\xi_2]\rangle + 
\langle a, [\xi_1,\eta_2]+[\eta_1,\xi_2]\rangle
\end{eqnarray*}
Similarly as above
$\xi-i\eta$ belongs to $\ker\Lambda_{1,0}$ if and only if
$[\xi,a]=0$, $[\eta,a]=0$, i.e.,
$$
\ker\Lambda_{1,0}=\ann_\mathfrak g(a)+i\ann_\mathfrak g(a).
$$
We need to find the dimension of the space
\begin{eqnarray*}
K &=&\{\xi-i\eta\in\ker\Lambda_{1,0}\,\vert\,\Lambda_{0,1}(\xi-i\eta,\ker\Lambda_{1,0})=0\}\\
&=&\{\xi-i\eta\in\ker\Lambda_{1,0}\,\vert\, 
(\epsilon a+ ia+ib,[\xi-i\eta,\ann_\mathfrak g(a)+i\ann_\mathfrak g(a)])=0\}\\
&=&\{\xi-i\eta\in\ker\Lambda_{1,0}\,\vert\, 
(ib,[\xi-i\eta,\ann_\mathfrak g(a)+i\ann_\mathfrak g(a)])=0\}
\end{eqnarray*}

Whence, $K$ consists of those elements in $\ann_\mathfrak g(a)+i\ann_\mathfrak g(a)$
which commute with $ib$. But, since $ib$ is a regular element of $\mathfrak g_0$, 
we find $\dim K=2r$. The theorem is proved. $\Box$

\begin{remark}{\rm
From Corollary \ref{complete1}, by taking $b=0$, we get complete 
integrability of the magnetic geodesic flows of normal metrics 
on regular orbits $\mathcal O(a)$. It is interesting that in this case we have
$$
\mathcal F_1^\epsilon=\Theta_\epsilon^*\mathfrak A,
$$ and $\Theta_\epsilon^*\mathfrak B$ coincides 
with the set of commuting $G$-invariant functions on $T^*\mathcal O(a)$ obtained by shifting of
argument (\ref{shift}):
$$
\mathcal B_a=\Theta_\epsilon^*\mathfrak B.
$$
In this sense, remarkably, the shifting
of argument method
\cite{MF1} can be seen as a particular case of the method of
contraction of Lie algebras \cite{Bo1, Br}.
By modifying the proof of Theorem \ref{complete2},  
it would be possible to give a proof of the completeness of 
$\mathcal B_a+\mathcal F_1^0$ on singular orbits $\mathcal O(a)$, 
different from those given in \cite{BJ3, MP}.
}\end{remark}

\begin{remark}{\rm 
As it follows from Reyman and Semenov-Tian-Shanski 
\cite{RS} (or by straightforward computation) 
the equations (\ref{x}), (\ref{phi}) are 
equivalent to the L-A pair
$$
\dot L(\lambda)=[L(\lambda),A(\lambda)] 
$$ 
with a spectral parameter $\lambda$, where 
$
L(\lambda)=
\lambda \Phi_\epsilon + 
i(x+\lambda^2  b),\,A(\lambda)=\Phi_\epsilon + i \lambda  b.
$ 
The integrals $\Theta_\epsilon^*\mathfrak B$ 
are exactly the integrals arising from the 
L-A representation of the system. }\end{remark}


\begin{thebibliography}{99} 

%\medskip 
\small

\bibitem{AM} Abraham, R. and  Marsden, J. E.: 
{\it Foundations of mechanics.} Second edition,  Benjamin/Cummings Publishing Co., Inc., Advanced 
Book Program, Reading, Mass. 1978 
 
\bibitem{Ba} Baguis, P.: Semidirect products and the Pukanszky condition. J. 
Geom. Phys. {\bf 25}, no. 3-4, 245--270 (1998)
 
\bibitem{BBC} Bloch, A. M, Brockett, R. W. and Crouch P. E.: 
Double Bracket Equations and Geodesic Flows on Symmetric Spaces. 
Commun. Math. Phys. {\bf 187}, 357-373 (1997) 
 
\bibitem{Bo1} Bolsinov, A.V.: 
Completely integrable systems on contractions of Lie algebras, 
Tr. Sem. Vekt. Tenz. An. XXII. M.: Izd-vo MGU, 8-16 (1985) (Russian)
 
\bibitem{Bo} Bolsinov, A. V.:   {Compatible Poisson brackets on Lie 
algebras and the completeness of families of functions in involution}, 
Izv. Acad. Nauk SSSR, Ser. matem. {\bf 55}, no.1, 68-92 (1991) (Russian); 
English translation: Math. USSR-Izv. {\bf 38},  no.1, 69-90  (1992) 
 
 \bibitem{BJ1} Bolsinov, A. V. and Jovanovi\' c, B.: 
{Integrable geodesic flows on homogeneous spa\-ces}. 
Matem. Sbornik {\bf 192}, no. 7, 21-40 (2001) (Russian); 
English translation: Sb. Mat. {\bf 192}, no. 7-8, 951-969 (2001)
 
\bibitem{BJ2} Bolsinov, A. V. and Jovanovi\' c, B.: 
{Non-commutative integrability, moment map and geodesic flows}.
Annals of Global Analysis and Geometry {\bf 23}, no. 4, 305-322 (2003), 
arXiv: math-ph/0109031
 
\bibitem{BJ3} Bolsinov, A. V. and Jovanovi\' c, B.: 
Complete involutive algebras of functions on 
cotangent bundles of homogeneous spaces. 
Mathematische Zeitschrift {\bf 246} no.  1-2, 213--236 (2004)

\bibitem{BJ4} Bolsinov, A. V. and Jovanovi\' c, B.: 
Magnetic Geodesic Flows on Coadjoint Orbits, to appear in J. Phys. A: Math. Gen.  {\bf 39}, L247--L252(2006), 
arXiv: math-ph/0602016

\bibitem{Br} Brailov, A. V.: Some constructions of the complete commutative sets of functions.
Tr. Sem. Vekt. Tenz. An. XXII. M.: Izd-vo MGU, 273-276 (1985) (Russian)
 
\bibitem{Br3} Brailov, A. V.: 
{Construction of complete integrable geodesic flows  on compact symmetric spaces}. 
Izv. Acad. Nauk SSSR, Ser. matem. {\bf 50},   no.2, 
661-674  (1986) (Russian); English translation: 
{Math. USSR-Izv.} {\bf 50}, no.4, 19-31 (1986)
 
\bibitem{Bul}  Buldaeva, E. A.: 
On integrable geodesic flows on the adjoint orbits  of orthogonal groups.
Diploma work, Dept. of Mathematics and Mechanics, 
Moscow State University, May 2002, (Russian)
 
\bibitem{DET} Duval, C., Elhadad, J. and Tuynman, G. M.: 
Pukanszky's condition and symplectic induction. J. Differential Geometry
{\bf 36}, 331-348 (1992)

\bibitem{Ef1} Efimov, D. I. The magnetic geodesic flows in a  homogeneous field on the 
complex projective space. Siberian Mathematical Journal {\bf 45}, no.3, 465-474 (2004)
 
\bibitem{Ef2} Efimov, D. I. The magnetic geodesic flows on a homogeneous symplectic manifold.
Siberian Mathematical Journal {\bf 46}, no.1, 83-93 (2005)
 
\bibitem{GS} Guillemin, V and Sternberg, S.: 
{\it Symplectic techniques in physics.} Cambrige University press, 1984

\bibitem{H}  Hedlund, G.A.: Fuchsian groups and transitive horocycles. 
 Duke Math. J. {\bf 2} , 530-542, (1936).

\bibitem{Jo} Jovanovi\' c, B.: 
On the Integrability of Geodesic Flows of Submersion Metrics. Lett. Math. Phys. 
{\bf 61}, 29-39 (2002), arXiv: math-ph/0204048 

\bibitem{KP} Kamalin, S. A. and Perelomov, A. M.: Construction of 
Canonical Coordinates on Polarized Coadjoint Orbits of Lie Grops.
Comm. Math. Phys. {\bf 97}, 553-568 (1985)
 
\bibitem{Ku} Kummer, M.: On the construction of the reduced
phase space of a Hamiltonian system with symmetry. Indiana Univ. Math. J. {\bf 30},
281-291 (1981)

\bibitem{Ma} Manakov, S. V.: Note on the integrability of the Euler equations
of $n$--dimensional rigid body dynamics, Funkc. Anal. Pril. {\bf 10}, no. 4, 93-94 (1976)
(Russian).

\bibitem{Mik1} Mikityuk, I. V.: 
{Homogeneous spaces with integrable $G$--invariant Hamiltonian flows}. 
Izv. Acad. Nauk SSSR, Ser. Mat. {\bf 47}, no.6, 1248-1262  (1983) (Russian)
 
\bibitem{MP} Mykytyuk, I. V. and Panasyuk A.: Bi-Poisson structures and integrability 
of geodesic flows on homogeneous spaces. Transformation Groups {\bf 9}, no. 3, 289-308 (2004)
 
\bibitem{MF1} Mishchenko, A. S. and Fomenko, A. T.: 
{Euler equations on finite-dimensional Lie groups}.
Izv. Acad. Nauk SSSR, Ser. matem. {\bf 42}, no.2,  396-415  (1978) 
(Russian);    English translation: Math. USSR-Izv. {\bf 12}, no.2, 
371-389  (1978)
 
\bibitem{MF2}  Mishchenko, A. S. and Fomenko, A. T.: 
{Generalized Liouville method of integration of Hamiltonian systems}.
Funkts. Anal. Prilozh. {\bf 12}, No.2, 46-56  (1978) (Russian); 
English translation: Funct. Anal. Appl. {\bf 12}, 113-121  (1978)
 
\bibitem{NS} Novikov, S.P. and Shmeltzer, I.: Periodic solutions 
to Kirchoff equations for a free motion of a rigid body in fluid 
and Lusternik-Shnirelman-Morse extended theory. Funct. Anal. Appl. 
{\bf 15}, 54-66 (1981)
 
\bibitem{N} Nekhoroshev, N. N.: Action-angle variables 
and their generalization. Tr. Mosk. Mat. O.-va. {\bf 26}, 181-198, (1972)  (Russian); 
English translation: {Trans. Mosc. Math. Soc.} {\bf 26},  180-198 (1972)

\bibitem{RatOrt}
Ortega, J.-P. and Ratiu, T. S.:
{\it Momentum maps and Hamiltonian reduction}.
Progress in Mathematics, 222.
Birkhuser Boston, Inc., Boston, MA, 2004. 


\bibitem{R} Reyman, A. G.: Integrable 
Hamiltonian systems connected with graded Lie algebras,
Zap. Nauchn. Semin. LOMI AN SSSR {\bf 95}, 3-54, (1980) (Russian);
English translation: J. Sov. Math. {\bf 19}, 1507-1545, (1982).

\bibitem{RS} 
Reyman, A. G. and Semenov-Tian-Shanski, M. A.: 
Group theoretical methods in the theory of finite dimensional integrable systems. 
In: Arnold, V. I., Novikov, S. P. (eds.) {\it Dynamical systems VII},  Springer 1994, pp. 116-225
 
\bibitem{Sa} Sadetov, S. T.: A proof of the Mishchenko-Fomenko conjecture 
(1981).  Dokl. Akad. Nauk {\bf 397}, no. 6, 751--754 (2004) (Russian)

\bibitem{Sak} Saksida, P.: Integrable anharmonic oscilators on spheres and hyperbolic spaces.
Nonlinearity {\bf 14}, 977-994 (2001)

\bibitem{Ta} Taimanov, A. I.:
An example of jump from chaos to integrability in magnetic geodesic flows. 
Matem. Zametki {\bf 76}, no. 4, 632-634 (2004) (Russian); English translation:
Math. Notes {\bf 76}, 587-589 (2004),  arXiv: math.DS/0312430

\bibitem{Th} Thimm A.: 
{Integrable geodesic flows on homogeneous spaces}. 
Ergod. Th. \& Dynam. Sys. {\bf 1}, 495-517  (1981)
 
\bibitem{Zu} Zung, N. T.: Torus actions and integrable systems.
In: {\it Topological methods in the theory of integrable systems}.  Bolsinov A.V., Fomenko A.T., Oshemkov A.A. (eds.)
Cambridge Scientific Publ., 2006,  arXive: math.DS/0407455
 
 
\end{thebibliography}
\end{document}